\documentclass[twocolumn,final]{svjour3}            
\usepackage{amsmath,amsfonts,amssymb,amscd,graphicx}

\newenvironment{Gauss}{\par\vskip1pt\noindent{\bf Gauss Principle.}\,\em}%
{\smallskip}

\newcommand{\nn}{\nonumber}

\def\vett#1{\underline{#1}}
\def\e_#1{\vett e_#1}
\def\k_#1{\vett k_#1}
\def\const{\text{cost.}}
\def\Re{\mathbb R}      

\def\a{\alpha}
\def\b{\beta}
\def\g{\gamma}
\def\muhat{\hat{\plus{5.5}0\mu\@}}
\def\nuhat{\hat{\plus{5.5}0\nu\@}}
\def\eps{\varepsilon}

\def\s{\sigma}

\def\v{\wedge}

\def\w{\omega}      
\def\x{\hat{\plus50 x\@}}

\def\xtilde{\raise-1ex\hbox to0pt{$\scriptstyle\sim\hss$} \@\?x{}}
\def\hxtilde{\raise-1ex\hbox to0pt{$\scriptstyle\sim\hss$} \@\x{}}
\def\B{\mathfrak B}
\def\D{\mathcal D}
\def\F{\mathfrak F}

\def\W{\Omega}

\def\de#1/de#2{\frac{\partial{#1}}{\partial{#2}}}
\def\d#1/d#2{\frac{d#1}{d#2}}
\def\sd#1/de#2/de#3{\ifx#2 \frac{\plus02\partial^{\@\@2}#1}{\plus90\partial\@#3^{\@2}}%
\else\frac{\plus02\partial^{\@\@2}#1}{\partial\?#2\@\partial\?#3}\fi}
\def\oD#1/d#2{\textstyle{\text{\large$\d{#1}/d{#2}$}}}
\def\De#1/de#2{\textstyle{\text{\large$\de{#1}/de{#2}$}}}
\def\Sd#1/de#2/de#3{\textstyle{\text{\large$\sd{#1}/de{#2}/de{#3}$}}}

\def\E{S\!\@\?E\/(3)}
\def\TE{T\!\@S\!\@\?E\/(3)}

\def\Tr#1{{\plus60}^t\!\?#1}

\def\@{\hskip.65pt}
\def\?{\hskip.3pt}
\def\plus#1#2{\vrule height#1pt width0pt depth#2pt}
\def\Tondo{\par\vskip2pt\noindent$\bullet\@$ }

\begin{document}

\title{Newton--Euler, Lagrange and Kirchhoff formulations of rigid body dynamics: a unified approach}

\author{Enrico Massa \and Stefano Vignolo}

\institute{E. Massa \at DIME, Sez.~Metodi e Modelli Matematici, Universit\`a di Ge\-no\-va. Piazzale Kennedy, Pad.~D. 16129 Genova (Italy).  \\\email{massa@dima.unige.it}
    \and
          S. Vignolo \at DIME, Sez.~Metodi e Modelli Matematici, Universit\`a di Ge\-no\-va. Piazzale Kennedy, Pad.~D. 16129 Genova (Italy). \\\email{vignolo@dime.unige.it}
}

\date{Received: date / Accepted: date}

\maketitle
\maketitle
\abstract{A unified formulation of rigid body dynamics based on Gauss principle is proposed. The Lagrange, Kirchhoff and Newton--Euler equations are seen to arise from
different choices of the quasi--coordinates in the velocity space. The group--theoretical aspects of the method are discussed.}
\keywords{Gauss principle \and Rigid body kinematics and dynamics \and Lagrange and Kirchhoff equations}
\PACS{45.40.-f \and 45.20.D \and 45.20.Jj}

\date{}

\section{Introduction}
Kirchhoff equations are a useful tool in rigid body dynamics: they are well known and widely used in many fields of applied mathematics, such as robotics, as well as
aerospace and naval engineering.

For example, they play a central role in the representation of the hydrodynamical forces acting on a rigid body moving in an incompressible, irrotational, and inviscid
fluid in terms of the so called \emph{added mass\/} \cite{Milne-Thomson,Newman}, an aspect that has significant applications in maneuvering models for surface vessels as
well as for underwater marine vehicles \cite{Fossen1,Fossen2,Lewandowski}.

The deduction of Kirchhoff equations from Newton--Euler ones is well known. On the contrary, the relationship between Kirchhoff and Lagrange equations is less immediate
and, in the authors' knowledge, not readily available in the literature.

In this brief note we fill this gap at the light of Gauss principle of least constraint.
The differences between the various formulations are shown to stem from different choices of the quasi coordinates in the velocity space.
In particular, Kirchhoff and Newton--Euler approaches are seen to reflect the Lie group structure of the configuration manifold associated with the free rigid body,
namely to involve quasi--coor\-dinates respectively associated with \emph{left--invariant\/} and \emph{right--invariant\/} $\@1$--forms.

Besides pointing out the centrality of Gauss principle, the present contribution is aimed at illustrating how the invariance properties of the configuration manifolds
help selecting families of independent variables especially suited to the formulation of the equations of motion. As such, it may provide a useful tool for possible
applications in different engineering disciplines.

\section{The euclidean group}
In this section, a few general aspects of rigid body kinematics are briefly reviewed.

Given a rigid body $\B$\vspace{1pt}, let $\@\F'=\big\{O,\e_1,\e_2,\e_3\big\}\@$ be a body--fixed cartesian (positive) reference frame, with coordinates
$\@x_i(P)=(P-O)\cdot\e_i=\const\@$  $\@\forall\;P\in\B\@$.\vspace{2pt}

Denoting by $\F=\big\{\W,\k_1,\k_2,\k_3\big\}$\vspace{1pt} a cartesian (positive) reference frame in the observer's space, with coordinates $\x_i(P)=(P-\W)\cdot\k_i\@$,
every configuration of $\B$ relative to $\F$ is described by the affine transformation
\begin{equation}\label{1.1}
\x_i(P)\@=\@R_{ij}\,x_j(P)\@+\@ b_i\,,\qquad i,j=1,2,3
\end{equation}
with $\@R_{ij}:=\k_i\cdot\e_j\@$ and $\@b_i:=x_i(O)=(O-\W)\cdot\k_i\@$.\vspace{2pt}

The totality of transformations \eqref{1.1} form a Lie group
\linebreak \cite{Warner,Sternberg}, henceforth denoted by  $\@\E\@$ and called the \emph{euclidean group\/}.

Introducing the column vectors $\hxtilde=\Tr\?\left(\x_1,\x_2,\x_3,1\right)$,\vspace{-.8pt} $\xtilde=\Tr\?\left(x_1,x_2,x_3,1\right)\@$,\vspace{.5pt} eq.\;\eqref{1.1}
takes the compact form
\begin{equation}\label{1.2}
\hxtilde\@=\@\begin{pmatrix}
R_{ij} & b_i\\[2pt]
0 & 1
\end{pmatrix}
\@\xtilde
\end{equation}
pointing out the isomorphism between $\@\E\@$ and the subgroup of $\@GL\?(4)\@$ given by the semi--direct product of $\@SO(3)\@$ with $\Re^3$.

In the given geometrical environment, every evolution of $\B$ is described by a curve $\g:\Re\to \E$, namely
\begin{equation}\label{1.3}
\g(t)\@=\@\begin{pmatrix}
R_{ij}(t) & b_i(t)\\[2pt]
0 & 1
\end{pmatrix}
\end{equation}

Denoting by $\hat u_i=\vett v_O\cdot\k_i$ and $u_i=\vett v_O\cdot\e_i$\vspace{1pt} the components --- respectively in the bases $\big\{\k_i\big\}$ and $\big\{\e_i\big\}$
--- of the velocity $\vett v_O=\oD/dt\Big|_{\F}(O-\W)$ of the body--fixed origin, we have the identifications
\begin{equation}\label{1.4}
\hat u_i=\dot b_i\,,\quad u_i = R_{ji}\,\dot b_j \quad\Longrightarrow\quad\hat u_i= R_{ij}\@u_j
\end{equation}

The angular velocity $\vett\w$ of $\B$ relative to the frame $\F$ is similarly expressed in components in either form
\begin{subequations}\label{1.5}
\begin{alignat}{3}
&\vett\w\@=\@\tfrac12\;\e_i\v\dot{\vett e}_i = &&\tfrac12\big(R\,\Tr{\dot R}\big)_{pq}\,\eps_{pqr}\@\k_r\@&&:=\@\hat\w_r\,\k_r          \label{1.5a} \\[4pt]
&\vett\w\@=\@\tfrac12\,\e_i\v\dot{\vett e}_i = &&\tfrac12\big(\@\Tr{\dot R}\@R\big)_{pq}\,\eps_{pqr}\@\e_r\@&&:=\@\w_r\,\e_r            \label{1.5b}
\end{alignat}
\end{subequations}
$\Tr{\dot R}$ and $\eps_{pqr}$ indicating the transpose of the matrix $\@\dot R\@$ and the permutation symbol.

\section{Group--invariant velocities}
As pointed out, the euclidean group $\E$ is the configuration space of the free rigid body. The tangent bundle $\TE$ is therefore identical to the associated velocity
space.

Every local coordinate system $q^\a$ ($\a=1,\ldots,6$) in $\E$ induces jet coordinates $q^\a\!\?,\dot q^\a$ in $\TE$. The latter are the ones commonly adopted in
Lagrangian Mechanics.

Other choices are of course available: for example, any $1$--form $\s=\s_\a\,dq^\a$ on $\E$ determines a corresponding function $\s_\a\@\dot q^\a$ that, under suitable
circumstances, can be adopted as fiber coordinate in $\TE$.

In this respect, the left and right invariant $1$--forms are naturally highlighted as distinguished geometrical objects, intimately related to the symmetries of the
underlying environment.

Any function on $\TE$ associated with a (left or right) invariant $1$--form $\@\s\in\D_1\/(\E)\@$ will be called a \emph{group--invariant velocity\/}.\vspace{1pt}

A left--invariant basis for the module $\@\D_1\/(\E)\@$ is given by
\begin{subequations}\label{2.1}
\begin{align}
&\mu_i= R_{ji}\,db_j                                                                                        \label{2.1a} \\[2pt]
&\nu_i=-\@\tfrac12\,\eps_{ipq}\,\big(\Tr R\,dR\big)_{pq}=-\@\tfrac12\,\eps_{ipq}\,R_{kp}\,dR_{kq}           \label{2.1b}
\end{align}
\end{subequations}

Denoting by $\@l_g:\E\to \E\@$ the left transport $\@l_g\/(h):=g\cdot h\@$, we have in fact the relation
\begin{equation*}
l^*_g\@\begin{pmatrix}
R_{ij} & b_i\\[4pt]
0 & 1
\end{pmatrix}\,=\,\begin{pmatrix}
R_{ip}\/(g)\@R_{pj}\quad &\quad R_{ij}\/(g)\,b_j\@+\@b_i\/(g)\\[4pt]
0 & 1
\end{pmatrix}
\end{equation*}
whence, by straightforward calculations
\begin{align*}
&l^*_g\left(\mu_i\right)= \hskip1ex R_{jp}(g)\@R_{pi}R_{jk}(g)\@db_k = \mu_i                                      \\[2pt]
&l^*_g\left(\nu_i\right)= -\@\tfrac12\,\eps_{ipq}\,R_{ks}(g)\@R_{sp}\@R_{kl}(g)\@dR_{lq} =\nu_i.
\end{align*}

The generalized velocities associated with the $1$--forms \eqref{2.1} are
\begin{subequations}\label{2.2}
\begin{align}
& \mu_i\quad\mapsto\quad R_{ji}\@\dot b_j\,=\, \vett v_O \cdot \e_i\,=\,u_i                                                                       \label{2.2a}\\[2pt]
& \nu_i\quad\mapsto\quad \hskip-1ex-\tfrac12\,\eps_{ipq}\,R_{kp}\@\dot{R}_{kq}\,=\,\w\cdot \e_i\,=\,\w_i                                          \label{2.2b}
\end{align}
\end{subequations}
i.e.~they coincide with the components, in the body--fixed basis, of the vectors $\vett v_O$ and $\vett\w$ involved in the representation
\begin{equation}\label{2.3}
\vett v_P\@=\@\vett v_O\,+\,\vett\w\v\big(P-O\big)\,, \qquad P\in \B
\end{equation}

A similar analysis shows that the $1$--forms
\begin{subequations}\label{2.4}
\begin{align}
&\muhat_i= db_i\@+\@R_{ip}\,dR_{jp}\,b_j                                                                                      \label{2.4a} \\[2pt]
&\nuhat_i=\@\tfrac12\,\eps_{ipq}\,\big(R\,d\,\Tr R\big)_{pq}=\@\tfrac12\,\eps_{ipq}\,R_{pk}\,dR_{qk}                         \label{2.4b}
\end{align}
\end{subequations}
form a right--invariant basis for the module $\@\D_1\/(\E)\@$.

Denoting by $\@r_g\/(h):=h\cdot g\@$ the right transport and arguing as above, we have in fact the relations
\begin{equation*}
r_g^*\big(R_{ij}\big)= R_{ip}\@R_{pj}\/(g)\,,\quad       r_g^*\big(b_i\big)=R_{ij}\,b_j\/(g)\@+\@b_i
\end{equation*}
whence
\begin{align*}
&r^*_g\left(\muhat_i\right)= db_i\@+\@R_{ip}\,dR_{jp}\,b_j  = \muhat_i                                            \\[2pt]
&r^*_g\left(\nuhat_i\right)= \@\tfrac12\,\eps_{ipq}\,R_{ps}\@R_{sk}\/(g)\@dR_{ql}\@R_{lk}\/(g) =\nuhat_i\@.
\end{align*}

The generalized velocities associated with the $1$--forms \eqref{2.4} are
\begin{subequations}\label{2.5}
\begin{align}
\muhat_i\quad\mapsto\quad &\dot b_i+\@R_{ip}\@\dot R_{jp}\,b_j\,=\,\dot b_i+\@\eps_{ijr}\@b_j\,\hat\w_p\,=           \nn         \\
&\,=\,\big(\vett v_O\@+\vett\w\v(\W-O)\big)\cdot\k_i\,:=\,\hat\xi_i                                                  \label{2.5a}\\[3.5pt]
\nuhat_i\quad\mapsto\quad &\tfrac12\,\eps_{ipq}\,R_{kp}\@\dot{R}_{kq}\,=\,\hat\w_i                                   \label{2.5b}
\end{align}
\end{subequations}
i.e.~they coincide with the components, in the observer's frame, of the vectors $\vett\xi$ and $\vett\w$ involved in the less usual representation\@
\footnote%
{According to eq.~\eqref{2.5a}, the vector $\@\vett\xi\@$ represents the velocity of the point of the body $\@\B\@$ instantly located at the space origin of the
observer's frame. As such, it may look a rather factitious object.
A~better understanding of the symmetry hidden in the representation \eqref{2.6} is gained interpreting the vectors $\@-\@\vett\xi$, $-\@\vett\w\@$ respectively as the
linear and angular velocity of the frame $\@\F\@$ relative to $\@\F'$, and the vector $\@-\@\vett v_P\@$ as the velocity, relative to $\@\F'$, of a point $\@P\@$ at rest
in $\@\F\@$. In this way, eq.~\eqref{2.6} is on the same footing as eq.~\eqref{2.3}, namely it describes, up to a sign, the rigid motion of $\@\F\@$ relative to
$\@\F'\@$. Interchanging left and right invariance is therefore equivalent to interchanging the roles of the frames $\@\F\@$ and $\@\F'$, i.e.~to replacing each
transformation by the corresponding inverse.}
\begin{equation}\label{2.6}
\vett v_P\@=\@\vett\xi\,+\,\vett\w\v\big(P-\W\big)
\end{equation}

\section{The equations of motion}
A central point in the development of mechanics in the presence of constraints is the characterization of the reactive forces. A milestone in this sense is provided by
the following
\begin{Gauss}
For a material system subject to ideal constraints, the actual motion under the action of given active forces is selected among the totality of kinematically admissible
evolutions by the requirement that, in any kinetic state $(t,P_1,\ldots,P_N,\vett v_1,\ldots,\vett v_N)$, the accelerations $\@\vett a_1,\ldots,\vett a_N\@$ are those
for which the function
\begin{equation}\label{3.1}
C:=\@\frac12\,\sum_{i=1}^N\,m_i\,\biggl|\vett a_i\@-\,\frac{\vett F_i}{m_i}\biggr|^2
\end{equation}
attains a minimum.
\end{Gauss}

For holonomic systems, Gauss' Principle is equivalent to d'Alembert's principle of \emph{virtual work\/} \cite{Levi,Whittaker,Hamel,mp1,mp2}.
The advantages of Gauss' formulation are its applicability to a wider class of constraints, including the kinetic ones \cite{mp1}, and its adaptedness to the language of
quasi--coordinates.

The implementation of the algorithm is straightforward: in terms of generic (fibred) coordinates $\@t,q^\a\!\?,z^\a$ on the velocity space, we have the relations
\begin{equation}\label{3.2}
\dot q^\a=\@\psi^\a\/(t,q^1\!\?,\ldots,q^n\!\?,z^1\!\?,\ldots,z^n)
\end{equation}
essentially equivalent to a definition of the generalized velocities, as well as the representations
\begin{align*}
 & P_i\,=\,P_i\/(t,q^1\!\?,\ldots,q^n)                                                                                               \\
 & \vett v_i\,=\,\de P_i/de{t}\,+\,\de P_i/de{q^\a}\,\psi^\a\@=\@\vett v_i\/(t,q^\a\!\?,z^\a)
\end{align*}

At each kinetic state, the expression of the admissible accelerations takes therefore the form
\begin{equation*}
\vett a_i\,=\de\vett v_i/de t\,+\,\de\vett v_i/de{q^\a}\,\psi^\a+\,\de\vett v_i/de{z^\a}\,\dot z^\a
\end{equation*}
involving $\@n\@$ additional variables $\@\dot z^\a$, inter\-pretable as coordinates along the fibres an affine bundle over the velocity space, known as the \emph{second
tangent bundle\/} \cite{Godb}.

It is then clear that imposing Gauss principle means, for each choice of $\@t,q^\a\!\?,z^\a\?$, minimizing the function \eqref{3.1} with respect to the variables $\@\dot
z^1,\ldots,\dot z^n$.\vspace{.5pt}
On account of the identity $\@\De\vett a_i/de{\dot z^\a}\@=\@\De\vett v_i/de{z^\a}\@$, this entails the condition
\begin{equation*}
\de\@C/de{\?\dot z^\a}=\sum_{i=1}^N\big(m_i\@\vett a_i-\vett F_i\big)\cdot\de\vett v_i/de{\?z^\a}\,=\,0
\end{equation*}
more conveniently written as
\begin{equation}\label{3.3}
\d/dt\biggl(\de\@T/de{z^\a}\biggr)-\@\sum_{i=1}^N\@m_i\@\vett v_i\,\d/dt\biggl(\de\vett v_i/de{\?z^\a}\biggr)=\,\sum_{i=1}^N\@\vett F_i\cdot\de\vett v_i/de{\?z^\a}
\end{equation}

Eqs.~\eqref{3.3} may be viewed as a set of equations for the determination of the unknowns $\@\dot z^\a\@$ in terms of the kinetic variables $\@t,q^\a\!,\?z^\a$
\footnote%
{Although conceptually preferable, for holonomic systems Gauss' principle is not strictly necessary in order to establish eq.~\eqref{3.3}: one may equally well start
with d'Alembert's principle, make use of the identity $\@\de P_i/de{q^\a}\@=\@\de\vett v_i/de{\dot q^\a}\@$, and replace the resulting equations by the linear
combinations
\begin{equation*}
0=\@\sum_{i=1}^N\big(m_i\@\vett a_i-\vett F_i\big)\cdot\de\vett v_i/de{\dot q^\b}\,\de\dot q^\b/de{\?z^\a}\,=\,
\sum_{i=1}^N\big(m_i\@\vett a_i-\vett F_i\big)\cdot\de\vett v_i/de{\?z^\a}
\end{equation*}
}.
We let the reader verify that the positive-definiteness of the matrix  $\@\Sd C/de\?\dot z^\a/de{\?\dot z^\b}=$ $=\sum_i m_i\,\De\vett v_i/de{z^\a}\cdot\De\vett
v_i/de{z^\b}\@$\vspace{2pt} ensures both the solvability of the equations and the fact that they do indeed determine a \emph{minimum\/} of the function $\@C\@$.

Summing up, we conclude that, eqs.~\eqref{3.3}, completed with the kinematical relations \eqref{3.2}, determine the evolution of the system from given initial data
through a well-posed Cauchy problem.

The covariance of the algorithm ensures that different choices of the generalized velocities lead to different but equivalent representations of the system \eqref{3.2},
\eqref{3.3}, without affecting the essence of the problem of motion, namely the determination of the curve $\@q^\a=q^\a\/(t)\@$ in configuration space.

As implicit in the notation, all previous results apply to \emph{discrete\/} systems.
In the continuous scheme, more suited to rigid body mechanics, the conclusions are essentially the same, with the concentrated attributes $\@m_i\@$, $\@\vett F_i\@$
replaced by corresponding \emph{measures\/} $\@d\/m\@$, $\@d\?\vett F\@$ over the abstract space $\@\B\@$ formed by the totality of points of the body, and with
eqs.~\eqref{3.3} replaced by the integral relations
\begin{equation}\label{3.4}
\d/dt\biggl(\de\@T/de{z^\a}\biggr)-\@\int_{\B}\@\vett v_P\,\d/dt\biggl(\de\vett v_P/de{\?z^\a}\biggr)\@d\/m\@=\int_{\B}\@\de\vett v_P/de{\?z^\a}\cdot d\?\vett F
\end{equation}

\smallskip
Coming to the problem in study, let us now verify that, depending on the choice of the generalized velocities, eqs.~\eqref{3.4} yield back the Lagrange, Kirchhoff and
Newton--Euler equations of motion.

\Tondo
{\em Jet coordinates:} with the ansatz $z^\a=\dot q^\a$, eqs.~\eqref{3.4} entail the Lagrange equations
\begin{equation*}
\d/dt\biggl(\de\@T/de{\dot q^\a}\biggr) -\,\de\@T/de{q^\a}\,=\int_\B\@\de\vett v_P/de{\dot q^\a}\cdot d\?\vett F\,=\int_\B\@\de P/de{q^\a}\cdot d\?\vett F
\end{equation*}
the right-hand sides expressing the so-called \emph{generalized forces\/}. The argument is well known, and does not require any comment.

\Tondo
{\em Left--invariant velocities:} as shown by eq.~\eqref{2.2},\vspace{.3pt} they correspond to the choice $z^i\!=u_i\@$, $z^{3+i}\!=\w_i\@$, $\@i=1,2,3\@$,
\mbox{$u_i=\vett v_O\cdot\e_i$} and $\@\w_i=\vett\w\cdot\e_i\@$ expressing the components of the linear and angular velocity of $\@\B\@$ in the body--fixed frame. On
account of eq.~\eqref{2.3}, this entails the identities
\begin{equation}\label{3.5}
\de\vett v_P/de{u_k} = \e_k, \qquad \de\vett v_P/de{\w_k} = \e_k \v (P- O)
\end{equation}

From eqs.~\eqref{3.5}, respectively denoting by $\vett Q$, $\vett\Gamma_0$, $\vett R$ and $\vett M_O$  the total linear momentum, the angular momentum (with respect to
$O$), the resultant of the external forces and the external torque relative to $\@O\@$, we have the identifications
\begin{subequations}\label{3.6}
\begin{align}
 &\de\@T/de{u_k}\@=\@\int_{\B}\@\vett v_P\,d\/m\cdot\e_k\@=\@\vett Q\cdot \e_k:=\@Q_k                                                       \label{3.6a}\\[5pt]
 & \de\@T/de{\w_k}\@=\@\int_{\B}\@\vett v_P\cdot\e_k \v (P- O)\@d\/m\@=\@\vett\Gamma_0\cdot\e_k:=\Gamma_k                                   \label{3.6b}\\[5pt]
 & \int_{\B}\@\de\vett v_P/de{u_k}\cdot d\?\vett F=\vett R\cdot \e_k:=\@R_k                                                                 \label{3.6c}\\[5pt]
 &\int_{\B}\@\de\vett v_P/de{\w_k}\cdot d\?\vett F=\vett M_O\cdot \e_k:=\@M_k                                                               \label{3.6d}
\end{align}
\end{subequations}

Inserting eqs.~\eqref{3.5}, \eqref{3.6} into \eqref{3.4} and recalling the Poisson formulae, we end up with the equations
\begin{align*}
& \d/dt\biggl(\de\@T/de{u_k}\biggr)+\@\eps_{kpq}\;\w_p\,\de\@T/de{u_q}\,=\,R_k                                                \\[3pt]
& \d/dt\biggl(\de\@T/de{\w_k}\biggr) +\@\eps_{kpq}\biggl(\w_p\,\de\@T/de{\w_q}\,+\@u_p\,\de\@T/de{u_q}\biggr)=\,M_k
\end{align*}
identical to the Kirchhoff equations.

\Tondo
{\em Right--invariant velocities:} according to eq.~\eqref{2.5}, they correspond to the ansatz $z^i\!=\hat\xi_i\@$, $z^{3+i}\!=\hat\w_i\@$,  $\@i=1,2,3\@$,
$\hat\xi_i=\vett\xi\cdot\k_i$ and $\@\hat\w_i=\vett\w\cdot\k_i\@$ being the components in the observer's frame of the vectors involved in the representation We have
therefore the identifications
\begin{equation}\label{3.7}
\de\vett v_P/de{\hat\xi_j} = \k_j, \qquad \de\vett v_P/de{\hat\w_j} = \k_j \v (P- \W)
\end{equation}

From these, preserving the notation $\@\vett Q\@$, $\@\vett R\@$ for the total momentum and total external force, and denoting by $\vett\Gamma_\W\,$, $\vett
M_\W$\vspace{.6pt} respectively the angular momentum and the torque with respect to the fixed origin $\W$, we get the relations
\begin{subequations}\label{3.8}
\begin{align}
& \de\@T/de{\hat\xi_s}=\vett Q\cdot\k_s:=\hat{\plus{7.5}0Q}_s\,,\qquad \de\@T/de{\hat\w_s} =\vett\Gamma_\W\cdot\k_s:=\hat\Gamma_s         \label{3.8a}\\[5pt]
& \int_{\B}\@\de\vett v_i/de{\hat\xi_s}\cdot d\?\vett F\,=\vett R\cdot\k_s:=\@\hat R_s                                        \label{3.8b}\\[5pt]
& \int_{\B}\@\de\vett v_i/de{\hat\w_s}\cdot d\?\vett F\,=\vett M_\W\cdot\k_s\@:=\@\hat M_s                                   \label{3.8c}
\end{align}
\end{subequations}

As above, inserting eqs.~\eqref{3.7}, \eqref{3.8} into \eqref{3.4} yields the equations
\begin{align*}
& \d\@\hat{\plus{7.5}0Q}_s/dt\,=\,\hat R_s                          \\[3pt]
& \d\@\hat\Gamma_s/dt\,=\,\hat M_s
\end{align*}
identical to the Newton--Euler equations.

\Tondo
{\em ``Hybrid'' formulation:} for completeness, we point out a further representation of the equations of motion,\vspace{.6pt} based on the choice $z^i=\dot
b_i\@,\,z^{3+i}=\w_i\@$,\vspace{.6pt} $i=1,2,3$, i.e.~involving the components of the velocity $\vett v_O$ in the observer's frame and the components of the angular
velocity $\vett\w$ in the body--fixed frame.

Referring to eq.~\eqref{2.3}, we have now the identities
\begin{equation*}
\de\vett v_P/de{\dot b_s} = \k_s\,, \qquad \de\vett v_P/de{\w_s} = \e_s \v (P- O)
\end{equation*}
whence, keeping the same notation as above
\begin{align*}
&\de\@T/de{\dot b_s}=\hat{\plus{7.5}0Q}_s\,,\qquad\; \de\@T/de{\w_s}=\Gamma_s\hskip1cm                                                    \\[5pt]
&\int_{\B}\@\de\vett v_P/de{\dot b_s}\cdot d\?\vett F\,=\hat R_s\,,\qquad \int_{\B}\@\de\vett v_P/de{\w_s}\cdot d\?\vett F\,=\@M_s
\end{align*}

Substituting into eqs.~\eqref{3.4} yields the required
\linebreak
equations, synthetically expressed in vector form as
\begin{align*}
& \d/dt\,\vett Q\;=\,\vett R                                                                                          \\[4pt]
& \d/dt\@\bigg|_{\F'}\,\vett\Gamma_O + \vett\w\v \vett\Gamma_O + \vett v_O\v\vett Q\,=\,\vett M_O
\end{align*}

These reproduce once again the content of the Newton--Euler equations, with the angular momentum and the torque now referred to the body--fixed point $O$.

\end{document}